\documentclass[aps,prb,reprint,showpacs,superscriptaddress,floatfix]{revtex4-1}
\usepackage{graphicx}
\usepackage{amsmath}
\usepackage{color}
\usepackage[hidelinks,bookmarks=false]{hyperref}

\begin{document}

\title{Local density of states study of a spin-orbit-coupling induced Mott insulator Sr$_2$IrO$_4$}

\author{Jixia Dai}
\email[]{daij@colorado.edu}
\affiliation{Department of Physics, University of Colorado at Boulder, Boulder, CO 80309}

\author{Eduardo Calleja}
\affiliation{Department of Physics, University of Colorado at Boulder, Boulder, CO 80309}

\author{Gang Cao}
\affiliation{Center for Advanced Materials, University of Kentucky, Lexington, Kentucky 40506}
\affiliation{Department of Physics and Astronomy, University of Kentucky, Lexington, Kentucky 40506}

\author{Kyle McElroy}
\email[]{kyle.mcelroy@colorado.edu}
\affiliation{Department of Physics, University of Colorado at Boulder, Boulder, CO 80309}

\begin{abstract}
We present scanning tunneling microscopy and spectroscopy experiments on the novel $J_\mathrm{eff}\!=\!1/2$ Mott insulator Sr$_2$IrO$_4$.
Local density of states (LDOS) measurements show an intrinsic insulating gap of 620 meV that is asymmetric about the Fermi level and is larger than previously reported values.
The size of this gap suggests that Sr$_2$IrO$_4$ is likely a Mott rather than Slater insulator.
In addition, we found a small number of native defects which create in-gap spectral weight.
Atomically resolved LDOS measurements on and off the defects shows that this energy gap is quite fragile.
Together the extended nature of the 5d electrons and poor screening of defects help explain the elusive nature of this gap.
\end{abstract}


\maketitle

Unlike localized 3d electrons and their resultant Mott insulating states \cite{Zaanen1985}, 4d and 5d electrons are more extended in space and thus should have both reduced on-site Coulomb repulsion (U) and enhanced energy bandwidths (W).
Because of this, many 4d systems \cite{Nakatsuji2000,Perry2006,Martins2011} are close to the borderline (U$\sim$W) of localization and one might expect that 5d systems will be metallic (U$<$W).
However, many iridium oxides \cite{Crawford1994,Cao1998,Singh2010,Okabe2011} have instead been found to be insulators.
Recently, it has been shown that for octahedrally coordinated $\mathrm{Ir^{4+}}$ ($\mathrm{5d^5}$) the large spin-orbit coupling (SOC), inherent to such a heavy element ($Z=77$), can dominate the $\mathrm{t_{2g}}$ orbitals and split them into a $J_\mathrm{eff}\!=\!1/2$ doublet and a $J_\mathrm{eff}\!=\!3/2$ quartet \cite{Kim2008}.
In this strong SOC limit, the quartet band is fully occupied and the conduction band becomes the half-filled $J_\mathrm{eff}\!=\!1/2$ band with a much smaller W. This means that the iridium oxides can be driven across the Mott transition borderline with a much smaller U compared to 3d materials.

One of these 5d materials, $\mathrm{Sr_2IrO_4}$, is of particular interest, since it is the first material shown to have this SOC-induced insulating state \cite{Kim2008,Kim2009}. In addition, it bears many similarities with the well-studied $\mathrm{La_2CuO_4}$ \cite{Kim2012}, a parent compound of the high-T$_\mathrm{c}$ superconductors, leading to an appealing possibility of superconductivity in $\mathrm{Sr_2IrO_4}$ with doping \cite{Wang2011,Watanabe2013}.
However, one of the similarities to $\mathrm{La_2CuO_4}$, which relies on $\mathrm{Sr_2IrO_4}$ being in the strong SOC limit so that only the $J_\mathrm{eff}\!=\!1/2$ band needs to be considered, becomes less certain when one realizes that the electron hopping energy ($t\approx0.3$ eV \cite{Wang2011,Watanabe2010}) is close to the SOC energy ($\zeta\approx0.5$ eV \cite{Montalti2006}).
The deviation from the strong SOC limit has been confirmed by the recent work by Haskel \emph{et al.} \cite{Haskel2012}
Furthermore, instead of the Mott type insulator, the magnetically originated Slater insulator has also been proposed \cite{Arita2012}, partially supported by time-resolved optical measurements \cite{Hsieh2012}, while more recent investigations have shown the possibility of an intermediate regime \cite{Watanabe2014}.

A key to identify the nature of this insulator is to answer the important questions of what is the size of the insulating gap, $\Delta$, and how does it compare to other energy scales (e.g. $t$ and $\zeta$).
Surprisingly, measurements of the gap size from different probes in $\mathrm{Sr_2IrO_4}$ vary widely.
Gaps as low as $\sim0.1$ eV have been reported by fitting the resistivity data with a thermal activation model \cite{Shimura1995,Ge2011} and by calculations using local density approximation (LDA) with SOC and U \cite{Kim2008,Jin2009}.
On the other hand, angle-resolved photoemission spectroscopy (ARPES) measurements do not agree with this value since the maximum of the valence band is already lower than $-$0.1 eV \cite{Kim2008,Wang2012}.
Optical conductivity and resonant inelastic x-ray scattering (RIXS) studies both yield $\Delta\leq0.4$ eV \cite{Kim2008,Kim2012,Moon2009}.
To resolve these discrepancies, it is highly desirable to measure the gap with an experiment that directly probes the free charge carriers.

We have carried out scanning tunneling microscopy (STM) and spectroscopy (STS) experiments on $\mathrm{Sr_2IrO_4}$.
Through single-particle tunneling, STS measurements cover both occupied and unoccupied states, yielding the local density of states (LDOS) to a good approximation and hence the single particle insulating gap directly \cite{Feenstra1988,Ye2013, Okada2013}.
Furthermore, the spatial resolution enables us to study variation down to the atomic-scale and to search for native defects that can pin the chemical potential and affect other bulk properties.
In this letter, we present high quality atomically resolved topographic \footnote{The topographic images have been processed to remove intermittent vibrational noise at 45 Hz.} and spectroscopic studies in $\mathrm{Sr_2IrO_4}$, uncovering a gap of 620 meV.
We have also studied topography with various junction resistances, and measured the LDOS with respect to the defects and to the tip-sample separation.
It is worth mentioning that similar STM studies have been done by other groups \cite{Nichols2014,Li2013a}, although the results and conclusions are not quite converging.

In this study, single-crystals of $\mathrm{Sr_2IrO_4}$ were cleaved in ultra-high vacuum (5$\times$$10^{\text{-}11}$ torr) immediately before being loaded into the STM scan-head, whose temperature is precisely stabilized at 80 K.
Before the tunneling experiments, we verified that the density of states of the tips were featureless.
For this data the junction was measured to have a work function of $3.6\pm0.2$ eV, which is necessary for measurements with bias voltages up to $\mathrm{\pm1}$~V.
Due to the insulating nature of $\mathrm{Sr_2IrO_4}$ 
measurements with high temperature and with large junction resistance ($>10$ $\mathrm{G}\Omega$) are necessary (see below) to only minimally perturb the surfaces. 
Furthermore, $\mathrm{Sr_2IrO_4}$ cleaves between two SrO layers leaving a charge-balanced surface, unlike the various surface terminations resulted from polar cleave such as the 122 Fe-based superconductor \cite{Zeljkovic2013}.

\begin{figure}[t]
    \includegraphics[width=3.25in]{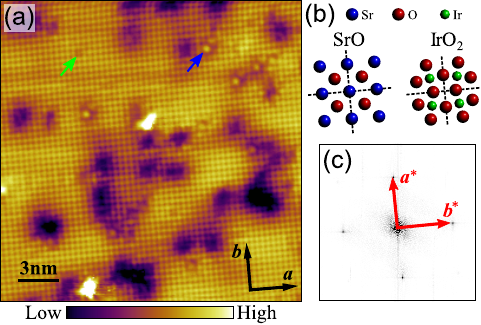}%
    \caption{\label{fig1}(Color online) A 20$\times$20 nm$^2$ topographic image taken with $-$300 mV and 5 pA at 80 K (a) and its Fourier transform (c). The black arrows indicate the $\boldsymbol{a}$ and $\boldsymbol{b}$ axes. Green (gray) and blue (dark) arrows indicate two Sr-related defects. (b) Schematic diagrams of the first SrO and second IrO$_2$ layers. The square lattice in (a) correspond to Sr.}
\end{figure}

After cleaved, $\mathrm{Sr_2IrO_4}$ shows a SrO layer on the surface.
Fig.~\ref{fig1}a shows a 20$\times$20 nm$^2$ constant-current image, in which we can easily see a square lattice (Fig.~\ref{fig1}b).
This surface lattice could be assigned as Sr, since when we partially replace Ir by Rh, the Rh dopants locate in-between four surface atoms \cite{Dai}.
The Fourier transform (Fig.~\ref{fig1}c) of this image shows four atomic peaks, confirming that this high quality surface is free of reconstruction.
Together with the atomic lattice, the sample also shows a small number of defects with the most evident type being the dark patches (see below).

\begin{figure}[b]
    \includegraphics[width=3.25in]{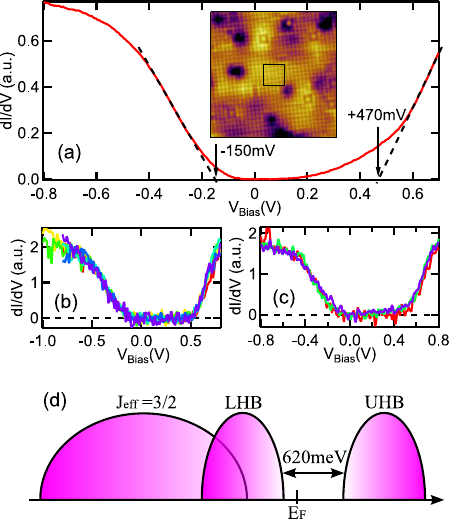}%
    \caption{\label{fig2}(Color online) The 620 meV energy gap. (a) LDOS measured by averaging 36$\times$36 spectra over the 2$\times$2 nm$^2$ area indicated by the square in the image (inset). The dashed lines are drawn to indicate the band edges at $-$150 mV and +470 mV. Data was taken with -300 mV, 10 pA and ac modulation of 8 mV$_\mathrm{rms}$. (b) and (c) LDOS taken at locations away from the defects, (b) without and (c) with the slow rise. (d) Diagram showing energy bands with two important features: the 620 meV insulating gap and the overlap between the LHB and the $J_\mathrm{eff}\!=\!3/2$ (quartet) band.}
\end{figure}

Shown in Fig.~\ref{fig2}a is an LDOS spectrum measured by averaging $\sim$1300 spectra taken over a 2$\times$2 nm$^2$ area that did not include any apparent defects.
The negative sample-bias side of the LDOS, corresponding to the occupied states, shows no distinct feature that can be related to the lower Hubbard band (LHB) of $J_\mathrm{eff}\!=\!1/2$ band down to $-$1 eV (Fig.~\ref{fig2}b).
This observation may seem to contrast with the double-peak structure observed in optical conductivity measurements \cite{Moon2009}.
However, these measurements in insulators are sensitive to excitons \cite{Kittel2004} and Kim \emph{et al.} \cite{Kim2012a} have pointed out that the double-peak seen in the optical conductivity are not due to the quartet band and the LHB.
Additionally, in the related compounds $\mathrm{Na_2IrO_3}$ and $\mathrm{Li_2IrO_3}$, the importance of excitonic effects have been shown in a recent RIXS study \cite{Gretarsson2013}.
It is therefore likely that the LHB is overlapping with rather than separate from the quartet band, in good agreement with calculations done by LDA+DMFT \cite{Arita2012} and by variational cluster approximation \cite{Watanabe2010}.
According to Watanabe \emph{et al.} \cite{Watanabe2010}, the LHB is entirely enclosed by the quartet band.
This overlap 
implies that the strong SOC limit and single-band models are insufficient for $\mathrm{Sr_2IrO_4}$. It is important to note that our data indicates that the SrO layer has little effect on the LDOS because defects like those shown by the arrows in  Fig.~\ref{fig1}a in this layer show no strong LDOS feature. 
In addition, any DOS contribution from this layer would give additional weight to our spectra and serve to make our gap measurement lower than the $\mathrm{IrO_2}$ layer. 
While ARPES hasn't identified the nature of all the measured bands their measurements agree that the only states within $-1$~eV on the filled side of the Fermi level are from a single band in agreement with theory \cite{Wang2012}. This means that if there is a Sr band it is mixed with the $\mathrm{IrO_2}$ ones and will not give a separate contribution to the LDOS.

\begin{figure}[b]
    \includegraphics[width=3.25in]{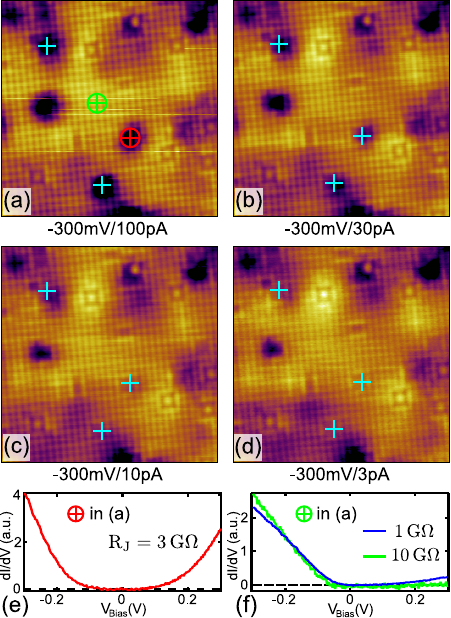}%
    \caption{\label{fig3}(Color online) (a-d) Images with a wide range of junction resistances: 3$-$100 G$\Omega$. The + signs indicate defects looking differently with tunneling condition. (e) LDOS measured on the red (dark) $\bigoplus$ in (a). (f) LDOS measured on the green (gray) $\bigoplus$ in (a) with 1/10 $\mathrm{G\Omega}$.}
\end{figure}

The insulating gap from our tunneling spectra is about 620 meV, with the valence band top at $-$150 meV and conduction band bottom at +470 meV (Fig.~\ref{fig2}a), which is much larger than the gap value reported by Li \emph{et al.}\cite{Li2013a}
Although similar gap size has been claimed by Nichols \emph{et al.} \cite{Nichols2014}, their definition of the gap is quite different from ours.
Here the energy gap is defined as the size of the energy window without density of states, and the edges of the gap are determined by linearly extrapolating the bands to vertical zeros \footnote{Due to the lack of a fine modeling of the band edges and the thermal broadening at 80 K, the exact gap size could even be larger than 620 meV and the uncertainty of our measurement could be as large as 30 meV.}, similar to the definition used by Okada \emph{et al.} in their study of $\mathrm{Sr_3Ir_2O_7}$ \cite{Okada2013}.
The energy of the valence band top ($-$150 meV) is in excellent agreement with the one measured by ARPES \cite{Kim2008,Wang2012}.
The slow rise in the LDOS between 0.2 and 0.5 eV varies with tunneling location and tip-induced electric field and hence we believe it to be extrinsic either due to local defect states or variable tip-induced band bending.
This extrinsic nature is further confirmed by the uniformity of the 620 meV gap unlike the slowly rising LDOS inside it which is seen only in some but not all of the locations (Figs.~\ref{fig2}b and \ref{fig2}c) and at different junction resistances (Fig.~\ref{fig3}f).
The reduction of this feature with decreased tip-induced electric field is consistent with it not being the intrinsic spectrum but instead due to band bending and defect related charging.
We report values for the large gap that are the limits as we lower the perturbing field.
Therefore, the large 620 meV gap is due to the underlying nonlocal energy bands. 
Furthermore, LDOS measurements in Rh-doped sample Sr$_2$Ir$_{1\text{-}\mathrm{x}}$Rh$_\mathrm{x}$O$_4$ \cite{Dai} also confirm this by showing that far from the Rh sites a similarly sized gap is present.

However, as discussed above, this large gap is not in good agreement with the gap values reported by other probes.
It is interesting to note that in the similar material $\mathrm{La_2CuO_4}$, the debate about the gap size has lasted for a long time \cite{Uchida1991,Kim2002,Ellis2008}.
The difficulty in extracting the correct energy scales of the intrinsic gap rises when different types of experiments actually probe them using different processes.
STS can measure the Mott gap via tunneling processes directly associated to the band edges.

Naturally, this 620 meV gap suggests that $\mathrm{Sr_2IrO_4}$ is unlikely a Slater insulator, since the magnetic coupling energy is merely 60$-$100 meV \cite{Kim2012,Fujiyama2012} which is too small to account for the large gap.
Additionally, while the long-range magnetic ordering temperature is $T_N$ = 240 K, resistivity measurement does not see a metallic state up to 600 K \cite{Chikara2009} suggesting that the magnetic ordering is unlikely the source of the insulating behavior.
The STS gap measurement should be further checked by experiment at temperatures above $T_N$, in which a reliable measurement has yet to be established. Thus, it still possible that the magnetic ordering is playing a role in the measured gap of 620 meV, such as in the recent proposal of intermediate Mott-Slater regime \cite{Watanabe2014} or in a strong coupling mechanism.

Although the size of this gap is considerably smaller than the cuprates \cite{Ye2013,Uchida1991}, it is still large compared to what one would expect from the extended 5d orbitals and the fact that $\mathrm{Sr_2RhO_4}$, which we naively expected to have a higher U and lower W, is actually a paramagnetic Fermi liquid \cite{Perry2006,Baumberger2006}.
The band minimum of the upper Hubbard band (UHB) of the $J_\mathrm{eff}\!=\!1/2$ band, is the major discrepancy between our STS result and the existing studies.
LDOS measurements show that the band minimum of UHB is at +470 meV, implying that LDA calculations need a larger U to capture this feature since it has been shown that the insulating gap depends on U sensitively \cite{Arita2012}.
In a recent study on $\mathrm{Na_2IrO_3}$, Comin \emph{et al.} \cite{Comin2012} have shown that using U as large as 3 eV in LDA calculation is necessary to reproduce the 340 meV gap in that material.
With this work a stronger-than-expected correlation effect has now been observed in two iridates.
The measured gap in $\mathrm{Sr_2IrO_4}$ is nearly twice the measured gap seen in $\mathrm{Na_2IrO_3}$, indicating that in the material studied here the 5d electrons are even more correlated (larger U/W).
Moreover, $\mathrm{Ba_2IrO_4}$ is also an insulator \cite{Singh2010} with a larger W due to the absence of the octahedra rotation and recent ARPES study has shown that its gap does not close above $T_N$ \cite{Moser2014}, which further confirms the strong entanglement of the energy scales in the 5d electrons.

The spatially extended nature of the 5d electrons would seem to greatly reduce the on-site Coulomb repulsion, but will concomitantly increase the same type of interactions between the electrons sitting on neighboring Ir sites due to the increased overlap of their wave functions.
This neighboring Coulomb interaction has been theoretically studied in the extended Hubbard model \cite{VandenBrink1995}.
For $\mathrm{Sr_2IrO_4}$, the Wannier functions calculated by Jin \emph{et al.} \cite{Jin2009} do show that a significant amount of electron weight is distributed on the four neighboring sites.
It is thus strongly possible that the symmetric arrangement of the neighbors can result in a net repulsion that acts as an effective on-site one.
Interestingly, every $\mathrm{Ir^{4+}}$ ion in $\mathrm{Sr_2IrO_4}$ has four nearest neighbors while there are only three in $\mathrm{Na_2IrO_3}$ which has a smaller gap.
This is in agreement with their gap sizes if the neighboring interaction is indeed causing a larger Hubbard-U.

To understand the effects caused by the defects, we have further studied topography at different junction resistances and LDOS on and off the defects.
We see two types of Sr-related defects in the topography: Sr vacancy (indicated by the green arrow in Fig.~\ref{fig1}a) and Sr ad-atom (blue arrow in Fig.~\ref{fig1}a).
These two types of defects may be caused by the cleaving process, but we find they have little effect on the local electronic structure, as discussed above.
In stark contrast the third type of defect we see, the larger dark patches seen in the topography, changes the local electronic structure dramatically.
We identify these dark-patch defects as related to oxygen defects, similar to those seen in manganites \cite{Bryant2011} and cuprates \cite{McElroy2005,Zeljkovic2012}.
Topographic images (Fig.~\ref{fig3}) with junction resistance ranging from 3.3 to 100 G$\Omega$ confirm that the patchiness associated with these defects are due to electronic inhomogeneity rather than surface structural corrugation.
The cross signs in Fig.~\ref{fig3}a-d indicate these oxygen defect related areas that vary with different tip-sample separation.
Such a high sensitivity in topography with respect to the tunneling junction reflects the poor screening in the presence of perturbations due to the insulating nature of $\mathrm{Sr_2IrO_4}$, and also the potential influence of the defects in these materials.
It is true that for some transition metal oxides, oxygen atoms in surface layers are volatile, but we didn't observe any degradation of surface quality after several days at 80 K implying the stability of the surface and these defects.
Therefore, these (O related) defects are likely native to the sample, in accord with the O deficiency found in this material \cite{Korneta2010,Qi2011}, and also in $\mathrm{Sr_3Ir_2O_7}$ and $\mathrm{BaIrO_3}$ \cite{Cao}.
This result naturally explains why transport measurements give a smaller gap \cite{Shimura1995,Ge2011} and optical conductance shows a slow rise starting at 0.2 eV \cite{Moon2009}.
The presence of these defects leads to in-gap states, lowers the effective gap measured by transport and gives a lower energy scale for optical transitions.
More importantly, the random distribution of the defects is in good agreement with the variable-range-hopping behavior in electrical conductivity \cite{Cao1998}.
The existence of native defects despite the good sample quality and our ability to measure the intrinsic LDOS apart from the influence of defects, prove the necessity of using a local probe like STM to study this material.

Lastly, the LDOS study on top of a defect (Fig.~\ref{fig3}e) shows how strongly the tip's electric field can change the density of states around $E_\mathrm{F}$ by building up in-gap states.
It should be pointed out that, due to the dramatic difference in topography of this defect as we mentioned earlier, the STS measurement may not be simply reflecting the underlying LDOS, but it does serve as evidence for the in-gap states around them.
Fig.~\ref{fig3}f shows the subtle variation in the measured spectra related to the tip-induced band bending, reinforcing the necessity of measuring with high junction resistances to reduce the perturbing electric field of the tip.

In summary, we have presented a high quality atomic-scale study in $\mathrm{Sr_2IrO_4}$ with density of states measurement covering the entire Mott gap, which is 620 meV.
The measurement of the occupied states has confirmed the overlap between LHB and quartet band.
We suggest that $\mathrm{Sr_2IrO_4}$ is a Mott but not Slater insulator after comparing the gap size with the magnetic coupling energy, implying the stronger-than-expected el-el correlation in this 5d system.
Furthermore, we propose that additional source of Coulomb repulsion needs to be taken into account to model the 5d electron systems.
Lastly, the junction dependence of the defects in the topography, combined with the LDOS results around them, indicate their profound impacts on the local electronic structure of this insulator.
Our results imply that in $\mathrm{Sr_2IrO_4}$ the three energy scales (SOC, U and W) are comparable to each other and that studies of the 5d electrons should put all of them on similar-footing.

\begin{acknowledgments}
Thanks to Michael Hermele, Gang Chen, Yue Cao and Dan Dessau for useful discussions.
G.C. acknowledges support by NSF through grants DMR-0856234 and EPS-0814194. K.M. acknowledges support by Alfred P. Sloan foundation.
\end{acknowledgments}


%

\end{document}